\documentclass[aps,prl,twocolumn,nopacs,superscriptaddress]{revtex4-1} 

\usepackage{amsmath}
\usepackage{color}
\usepackage{graphicx}

\begin{document}

\title{Measures of Quantum Synchronization in Continuous Variable Systems}

\author{A.\ Mari}
\affiliation{NEST, Scuola Normale Superiore and Istituto Nanoscienze-CNR, I-56127 Pisa, Italy}

\author{A.\ Farace}
\affiliation{NEST, Scuola Normale Superiore and Istituto Nanoscienze-CNR, I-56127 Pisa, Italy}

\author{N.\ Didier}
\affiliation{D\'epartement de Physique, Universit\'e de Sherbrooke, Sherbrooke, Qu\'ebec J1K 2R1, Canada}
\affiliation{Department of Physics, McGill University, Montreal, Quebec H3A 2T8, Canada}

\author{V.\ Giovannetti}
\affiliation{NEST, Scuola Normale Superiore and Istituto Nanoscienze-CNR, I-56127 Pisa, Italy}

\author{R.\ Fazio}
\affiliation{NEST, Scuola Normale Superiore and Istituto Nanoscienze-CNR, I-56127 Pisa, Italy}

\begin{abstract}
We introduce and characterize two different measures  which quantify the level of  synchronization 
of coupled continuous variable quantum systems.  The two measures allow to extend to the quantum domain 
the notions of complete and phase  synchronization. The Heisenberg principle sets a universal bound to complete 
synchronization. The measure  of phase synchronization is in principle unbounded, however in the absence of 
quantum resources ({\it e.g.}\ squeezing) the synchronization level is bounded below a certain threshold. We elucidate some 
interesting connections between entanglement and synchronization and, finally, discuss an application based on 
quantum opto-mechanical systems.
\end{abstract}

\maketitle

\begin{figure}[t]
 (a) \phantom{777777777777} (b) \phantom{77777777777777777777777}  \\
\vspace{- 0.01 cm}
\includegraphics[width=0.24 \columnwidth]{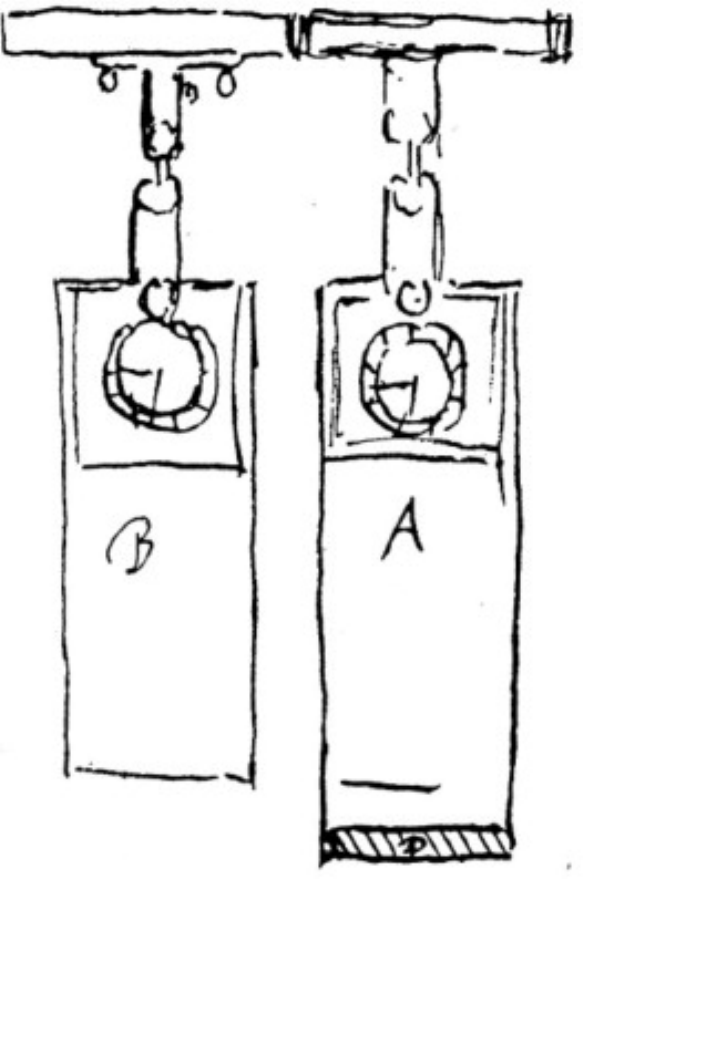} \includegraphics[width=0.63\columnwidth]{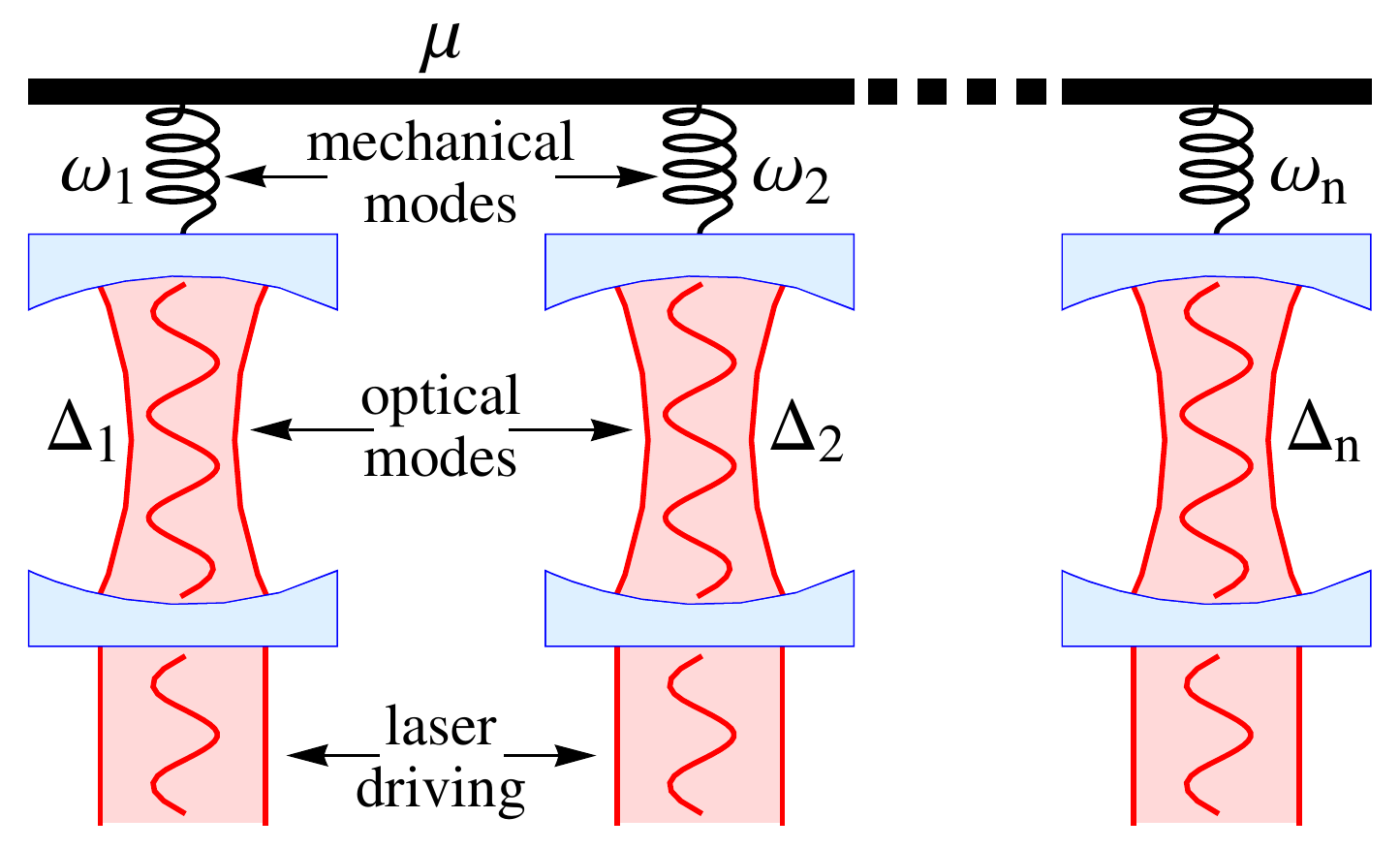} 
\caption{(Color online) Original Huygens' sketch \cite{huygens} of two synchronizing pendulum clocks (a) and the quantum mechanical analogue consisting of  two (or more) coupled opto-mechanical systems (b). Here, mechanical resonators are driven into self-sustained oscillations by the non-linear radiation pressure force of independent optical modes.  A weak mechanical interaction is responsible for the spontaneous synchronization of the limit cycles. All symbols are defined in the main text. 
}\label{model}
\end{figure}

In the 17th century, C. Huygens noticed that the oscillations of two pendulum clocks  with a common support tend to 
synchronize (Fig.\ \ref{model}.a)~\cite{huygens}. Since then, analogous phenomena have been observed in a large variety of different 
contexts,  {\it e.g.}\ neuron networks, chemical reactions, heart cells, fireflies, {\it etc.}~\cite{csync}. They are all instances of 
what it is called the {\it spontaneous  synchronization effect}   where  two or more systems, in the complete absence of 
any external time-dependent driving force,  tend to synchronize their motion solely due to their mutual coupling.
The emergence of spontaneous  synchronization in so many different physical settings encouraged its investigation 
within classical non-linear dynamical systems. Here, given the time evolution of two  dynamical variables, 
like the position  of two pendula, standard methods exist to verify whether their motion is synchronized~\cite{csync}.
For quantum systems however the same approaches cannot be straightforwardly extended due to the absence of a 
clear notion of  phase space trajectories.  The aim of this work  is to address this problem,  developing a consistent 
and quantitative theory of synchronization for continuous variable (CV) systems evolving in the quantum 
regime~\cite{CVSYSTEMS}.  To this aim we introduce two different quantum measures of synchronization extrapolating 
them from notions of complete and phase synchronization introduced for classical models.  We will show that quantum 
mechanics set bounds on the achievable level of synchronization between two CV systems and we will 
discuss the relationship between entanglement and synchronization.
We finally apply our approach for studying the dynamics of coupled opto-mechanical systems \cite{review1,review2}.

In the quantum domain  synchronization has been studied in various contexts, like
quantum information protocols 
\cite{protsyn}, two-level systems  \cite{quibit} and stochastic systems \cite{stochastic}. While our measures could in principle be extended also to these cases,
our endeavor  is specifically  framed in  the research line investigating the spontaneous synchronization of micro- and nano-mechanical systems \cite{array6,array4,array5,sync2,lee,sync3,sync4,sync5,matheny}. Recent experimental advances  allow to realize opto-mechanical
arrays composed  of two or more coupled mechanical resonators controlled close to their quantum regime by laser driving 
\cite{array1,array7,array8,array9}. Such devices have all the properties (non-linear dynamics, limit cycles, {\it
etc.})  which are necessary for the emergence of spontaneous synchronization \cite{array2,array4} and indeed some first experimental 
evidences of this effect have been found \cite{sync4, sync5, matheny}. 

{\it Quantum synchronization measures:--} In  a purely classical  setting, synchronization is mostly studied in the context of autonomous
non-linear systems undergoing limit cycles or chaotic evolution  (linear systems being usually excluded because they 
converge to constant or unstable solutions). In this scenario one  can identify  different forms of synchronization~\cite{csync}.
{\it Complete synchronization} is achieved  when (say) two subsystems $S_1$ and $S_2$, 
initialized into independent configurations, acquire identical trajectories  under the effects of  mutual interactions.
Specifically, given two CV classical systems characterized by the (dimensionless) canonical variables $q_1(t),p_1(t)$ and $q_2(t), p_2(t)$
describing the evolution of $S_1$ and $S_2$  in phase space, complete synchronization is reached when the quantities  $q_-(t):=[q_1(t)-q_2(t)]/\sqrt{2}$ and 
$p_-(t):= [p_1(t)-p_2(t)]/\sqrt{2}$ asymptotically vanish for large enough times \cite{factor}.
{\it Phase synchronization} is instead achieved when, under the same conditions detailed above,
only the phases $\varphi_j(t)= \arctan [p_j(t)/q_j(t)]$ are locked: {\it i.e.}\ when the quantity $\varphi_-(t) := \varphi_1(t)-\varphi_2(t)$ asymptotically converges to a constant phase shift $\varphi_0 \in [0,2\pi]$.

One can already foresee that extending the above concepts to quantum mechanical systems is not straightforward and 
that some fundamental limits could exist that prevent the exact fulfillment of the conditions given above. 
In particular, identifying the dimensionless quantities $q_j(t)$, $p_j(t)$  
as quadrature operators obeying the canonical commutation rules  $[q_j(t),p_{j'}(t)]=i\delta_{jj'}$ \cite{CVSYSTEMS}, the relative coordinates $q_-(t)$ and $p_-(t)$ 
will correspond to  generalized  position and momentum operators  of the same (anti-symmetric) mode of the system. Accordingly 
the uncertainty principle will now prevent the possibility of exactly achieving the condition required by classical complete synchronization.
  
To turn this into a quantitative statement, we identify $q_-(t)$ and  $p_-(t)$
as synchronization errors and introduce the following figure of merit
  \begin{eqnarray}
\mathcal S_{c}(t) &:=&\langle q_-(t)^2 + p_-(t)^2 \rangle ^{-1}, \label{PRIMA}
\end{eqnarray}
gauging the level of {\it quantum complete synchronization} attained by the system
(here $\langle \cdots\rangle$ implies taking the expectation value with respect to the density matrix of the quantum system).
We then observe that  the Heisenberg principle requires $\langle  q_-(t)^2\rangle \langle p_-(t)^2 \rangle \geq 1/4$  and hence
\begin{eqnarray}
\mathcal S_{c}(t) \le \frac{1}{2 \sqrt{\langle  q_-(t)^2 \rangle  \langle  p_-(t)^2 \rangle  } }\le 1,\label{heisenberg}
\end{eqnarray} 
 which sets a universal limit to the  complete synchronization two CV systems can reach.  On the contrary, in a purely classical 
 theory,  $\mathcal S_{c}(t)$ is in principle unbounded~\cite{NOTE1}. Indeed in real units 
 the right-hand side of the bound scales as $\hbar^{-1}$, diverging in the limit $\hbar \rightarrow 0$.
 
 A small value of $\mathcal S_{c}(t)$ can have two  
 possible origins:  the mean values of $q_-(t)$ and $p_-(t)$ are not exactly zero, and/or  the variances of such operators are large. The former situation can be interpreted as a classical systematic error \cite{error}, while the latter is due to the influence of thermal and quantum noise. The classical systematic error can be easily excluded from the measure of synchronization by using the same expression of Eq.\ (\ref{PRIMA}) but after the application of the 
change of variables 
\begin{equation}
q_-(t) \rightarrow q_-(t)- \langle q_-(t) \rangle, \quad p_-(t) \rightarrow p_-(t)- \langle p_-(t) \rangle \; . \label{relative}
\end{equation}
This gives a  relative measure of synchronization which is always larger than the previous absolute one and which may be preferable whenever the aim is that of selectively investigating purely quantum mechanical effects. Obviously, the bound of Eq.\ (\ref{heisenberg})  holds also for this relative  measure.

Constructing a quantum analogue of the phase synchronization condition  is more demanding due to the controversial 
nature of the quantum phase operator(s), see {\it e.g.}\ Ref.~\cite{PHASE}. In principle one could use a phase-difference operator as the one proposed in \cite{phase-diff},
however we adopt a more pragmatic approach which allows us to target departures from the ideal (classical) synchronization condition,  due to quantum fluctuations.
 To do so, we write the operator $a_j(t):=[q_j(t)+ i p_j(t) ]/ \sqrt{2}$ of the $j$-th system as 
\begin{equation}
a_j(t)=[ r_j(t) +a_j'(t) ] e^{i \varphi_j(t)}, \label{frame}
\end{equation}
where $r_j(t)$ and $\varphi_j(t)$ are the amplitude and phase of the expectation value of  $a_j(t)$,
 {\it i.e.}\ $\langle a_j(t) \rangle = r_j(t) e^{i \varphi_j(t)}$.
With this choice, the Hermitian and anti-Hermitian part of $a_j'(t)=[q_j'(t)+ i p_j'(t) ]/ \sqrt{2}$ can now be interpreted
as fluctuations of the amplitude and of the phase respectively (indeed this is the reason why in 
quantum optics $q'_j(t)$ and $p'_j(t)$ are often called {\it amplitude} and {\it phase} quadratures).
If two CV systems are on average synchronized such that the phases of $\langle a_1(t) \rangle $ 
and of $\langle a_2(t) \rangle $ are locked,  then the phase shift with respect to this locking 
condition can be associated to the operator
$p'_-(t)=[{p'_1(t)-p'_2(t)}]/{\sqrt{2}}$.
A  measure of {\it quantum phase synchronization} can  then be obtained through the quantity 
\begin{equation}
	\mathcal S_{p}(t):=\frac{1}{2}\langle p'_-(t)^2 \rangle^{-1}. \label{Spq}
\end{equation}
Differently from the measure (\ref{PRIMA}),
$\mathcal S_{p}$ can be in principle arbitrarily large. Nonetheless, 
if two CV quantum systems evolve in time such that their $P$-function \cite{glauber, CVSYSTEMS} is always positive (quantum optics
notion of classicality),  then perfect phase synchronization 
is impossible and one has 
\begin{equation}
	\textrm{positive}\  P \textrm{-function }  \Rightarrow   \mathcal S_{p}(t)  \le 1 \: . \label{bound2}
\end{equation}
Indeed a value of  $\langle p'_-(t)^2 \rangle $ below $1/2$ implies the existence of collective squeezing
and so the impossibility of a phase space representation of the state through a positive P-function. Notice that, with respect to the fundamental bound (\ref{heisenberg}), the threshold (\ref{bound2}) is much weaker since it can be overcome with squeezed states.

Furthermore the specific structure of the limit cycles associated with the average quantities $r_j(t)$ and $\varphi_j(t)$
may lead to additional bounds for $\mathcal S_{p}$.  If, for example,  {\it  i)}  the system under consideration exhibits  mean values quantities $\langle a_j(t)\rangle$ converging to approximately circular limit cycles in the phase space, {\it  ii)} the noise operating in the system is not phase sensitive 
({\it i.e.} is invariant for phase space rotations) and  {\it  iii)}  the interaction potential between the two systems is of the
form $H_{int}= -\mu (a_1 a_2^\dag+a_2 a_1^\dag)$; then it is reasonable to conjecture that $\langle p'_-(t) ^2\rangle \ge \langle q'_- (t)^2\rangle$.
This together with the  Heisenberg principle, leads to the bound
\begin{equation} 
\mathcal S_{p}(t)  \le \mathcal S_{c}(t)  \le 1 \;.
\label{conjecture}
\end{equation}
 While referring  to the Supplemental Material
for an heuristic derivation of Eq.~(\ref{conjecture}), we  remark that such inequality is consistent with the results
 shown later on opto-mechanical systems.

 {\it Quantum correlations and synchronization: --} 
Synchronization and entanglement are both associated with the presence of correlations between two or more systems.
It is thus natural to ask if, in the quantum regime, there is a strong interplay between the two effects.
 Quite surprisingly however it turns out that, according to our measures, the stationary state of two CV  systems can possess maximum amount of complete or phase synchronization {\it without}  being  necessarily entangled.  For instance a system converging to two factorized coherent states evolving in time such that $\langle a_1(t) \rangle=\langle a_2(t) \rangle$,  
exhibits  maximum complete synchronization ($S_c=1$) but  has  no entanglement. Similarly 
consider two locally squeezed states rotating in phase space such that $\langle a_1(t) \rangle=\langle a_2(t) \rangle$
and  $\langle p_1'(t)^2\rangle=\langle p_2'(t)^2\rangle=\epsilon $, with $p_k'$ being the quadrature orthogonal to the phase
space cycle of subsystem $k$ as defined in Eq.\ (\ref{frame}) (said in simpler words, these are two squeezed states 
moving like synchronized clock hands in phase space).  This state has arbitrary high phase synchronization 
$S_p=\frac{1}{2}\epsilon^{-1}$ but it is clearly not entangled. 
Entanglement appears hence to enforce  correlations which are qualitatively different from those required to
yield high values for ${\cal S}_c(t)$ and ${\cal S}_p(t)$. A better insight on this can be obtained  by considering 
 the very precursor  of all CV entangled states, {\it i.e.}\ the EPR state~\cite{EPR} which 
 describes the ideal scenario of  two systems  having  {\it same} positions but {\it opposite} momenta.
It is thus clear that synchronization requires different constraints which could have instead a relationship with other measures of quantum correlations like {\it quantum discord} (see {\it e.g.}\ our successive results on opto-mechanical systems).
We conclude this section with an open question on the converse problem: {what kind of synchronization phenomenon
 corresponds in the quantum limit to EPR correlations?}
EPR entanglement could be identified as a mixture of complete and anti-synchronization, {\it i.e.}\ 
$q_1(t)=q_2(t)$ and  
$p_1(t)=-p_2(t)$.
Recently this unconventional regime called {\it mixed synchronization} has been 
introduced and observed in classical non-linear systems  \cite{mixedsync1}, but whether 
this concept is relevant and extendible in the quantum domain is still unexplored.

{\it Measures and bounds at work:--}  Opto-mechanical devices~\cite{review2,review1}  provide the perfect setting where 
our measures for synchronization can be directly applied.  We thus identify $S_1$ and $S_2$   with two approximately identical mechanical resonators  (see Fig.~\ref{model}.b)
coupled to independent cavity optical modes (needed to induce self-sustained limit cycles)  and mutually 
interacting through a phonon tunneling term~\cite{array4}  of intensity $\mu$:
\begin{eqnarray}
H&=& \sum_{j=1,2} [- \Delta_j a_j^\dag a_j + \omega_j b_j^\dag b_j - g a_j^\dag a_j (b_j+b_j^\dag) \nonumber \\
&& + i E (a_j-a_j^\dag)]  - \mu (b_1 b_2^\dag+b_2^\dag b_1).  \label{HAM}
\end{eqnarray}
In this expression, for $j=1,2$, $a_j$ and $b_j$ are the optical and mechanical annihilation operators, $\omega_j$ are 
the mechanical frequencies, $\Delta_j$ are the optical detunings, $g$ is the opto-mechanical coupling constant, while $E$ is the laser intensity which drives the optical cavities ($\hbar=1$).
For simplicity $g$ and $E$ are assumed to be equal in both systems while $\omega_1$ and $\omega_2$ can be slightly different.
Dissipative effects are included adopting the Heisenberg picture and writing the following
quantum Langevin equations \cite{linearization},
\begin{eqnarray} \label{equ}
\dot{a}_j & =& [-\kappa + i \Delta_j +i g (b_j+b_j^\dag) ] a_j + E+\sqrt{2\kappa}a_j^{in} , \\
\dot{b}_j &= & [-\gamma - i \omega_j ] b_j +i g a_j^\dag a_j +i \mu b_{3-j}+\sqrt{2\gamma}b_j^{in} . \nonumber
\end{eqnarray}
Here $\kappa$ and  $\gamma$ are, respectively, the optical and mechanical damping rates  while  $a_j^{in}$ 
and $b_j^{in}$ are the input bath operators. These are assumed to be white Gaussian fields obeying standard correlation relations,  
$\langle a_j^{in} (t)^\dag a_{j'}^{in} (t') +a_{j'}^{in}  (t') a_j^{in} (t)^\dag \rangle = \delta_{jj'} \delta(t-t')$ and 
$\langle b_j^{in} (t)^\dag b_{j'}^{in} (t') +b_{j'}^{in}  (t') b_j^{in} (t)^\dag \rangle  = (2 n_b  +1) \delta_{jj'} \delta(t-t')$, 
where $n_b= [\exp(\frac{\hbar \omega_j}{k_B T})-1]^{-1}$
 is the mean occupation number of the mechanical baths which gauges the temperature $T$ of the system \cite{linearization}
(since we are only interested in the situation in which $\omega_1 \simeq \omega_2$, the parameter $n_b$ can be safely taken to be equal for both oscillators).

The operators $O(t)$ in Eq.~(\ref{equ}) can be expressed
 as sums of mean values $\langle O(t) \rangle$ plus fluctuation terms $O'(t)$, {\it i.e.}\ we write $O(t)=\langle O(t) \rangle+O'(t)$.
In a semiclassical approximation \cite{linearization} we determine the expectation values $\langle O(t) \rangle$ in terms of a set of classical non-linear differential equations and, as a second step, we linearize the quantum Langevin equations for the operators $O'(t)$. 
Setting $\Delta_j= \omega_j$ (driving detuning) and choosing the laser amplitude $E$ of Eq.~(\ref{HAM})  large enough, 
we make sure that such solutions yield limit cycles as classical steady state configurations
 (see {\it e.g.}\ \cite{limit-cycles}). In this regime the mechanical and optical fileds acquire large coherent amplitudes
 and therefore we expect the linearization procedure to be justified. A more general and exact treatment of the non-linear dynamics could be achieved by using stochastic methods like those presented in Ref.s \cite{P1,P2}.
 
   Quantum fluctuations  are obtained  by computing 
the  covariance matrix $C(t)$, with entries given by $C_{i,\ell}(t)=\langle R_i(t) R_{\ell}(t)^\dag +R_{\ell}(t)^\dag R_i(t) \rangle/2$, 
the expectation value being taken on the initial state and $R_i$ being the components of 
the vector  $R=(a_1',a_1'^\dag,b_1',b_1'^\dag,a_2',a_2'^\dag,b_2',b_2'^\dag)$.  
In particular this gives us direct access to the mechanical variances 
$\langle  q_-(t)^2 \rangle$ and $\langle  p_{-}(t)^2 \rangle$ which define the 
complete synchronization level via~Eq.(\ref{PRIMA}).  By applying the linearization procedure, we implicitly performed the change of variables corresponding to Eq.\ (\ref{relative}) and so we automatically excluded the systematic synchronization error due to slightly  different average trajectories. As a consequence the only source of disturbance bounding our measure of synchronization will be quantum (or thermal) fluctuations.   

Estimating  phase synchronization as in Eq.(\ref{Spq}) requires instead a 
further step as the latter has been defined with respect to a reference frame rotating with the phases of the average trajectories,
see Eq.\ (\ref{frame}). This corresponds to a diagonal and unitary operation on $R$,
built up on the phases $\varphi_{a_1}(t)={\rm arg}\langle a_1(t) \rangle $,  $\varphi_{a_2}(t)={\rm arg} \langle a_2(t) \rangle$, {\it etc.}, 
of the classical orbits: {\it i.e.}\ 
$R\rightarrow R'=U(t) R$ with 
$U(t)={\rm diag}[ e^{-i \varphi_{a_1}(t)}, e^{i \varphi_{a_1}(t)}, \cdots]$.
The associated covariance  matrix is $C'(t)=U(t) C(t) U(t)^\dag$, from which we can directly 
extract the mechanical variance $\langle p_-'^2 (t)\rangle$ entering  Eq.(\ref{Spq}).

\begin{figure}[t]
\includegraphics[width=0.49\columnwidth]{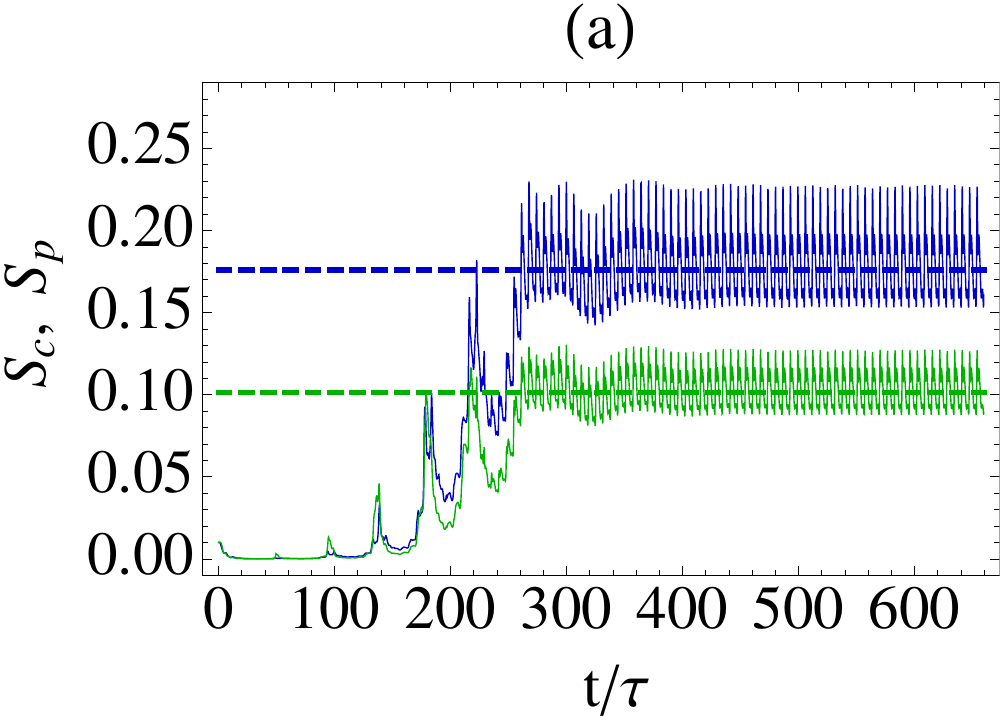} \includegraphics[width=0.49\columnwidth]{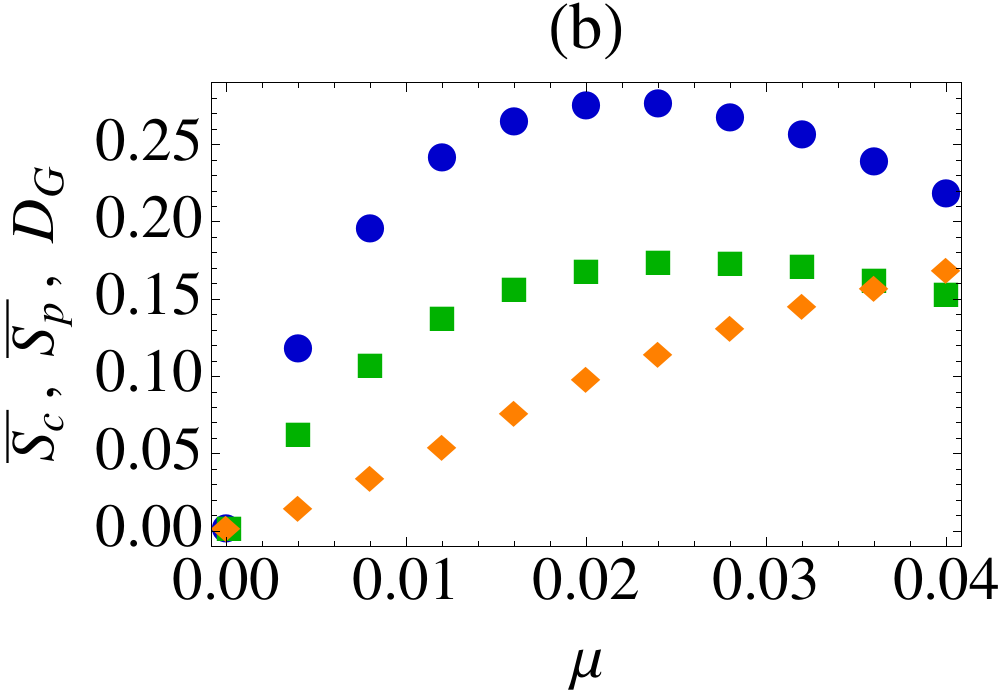} \\
\vspace{0.2 cm}
\includegraphics[width=0.49\columnwidth]{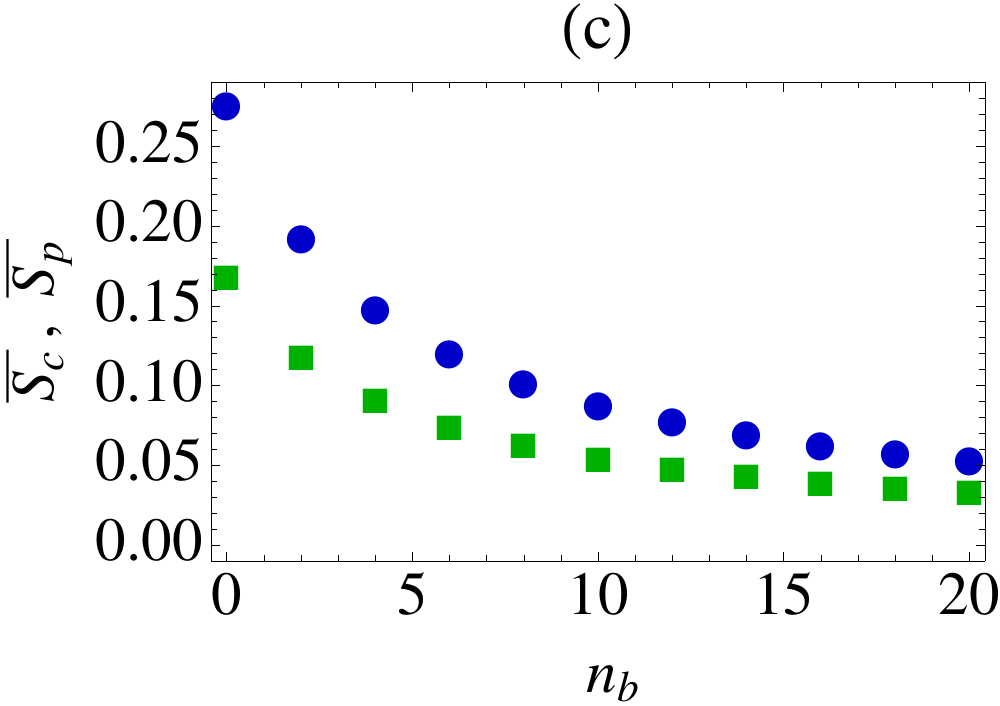} \includegraphics[width=0.49\columnwidth]{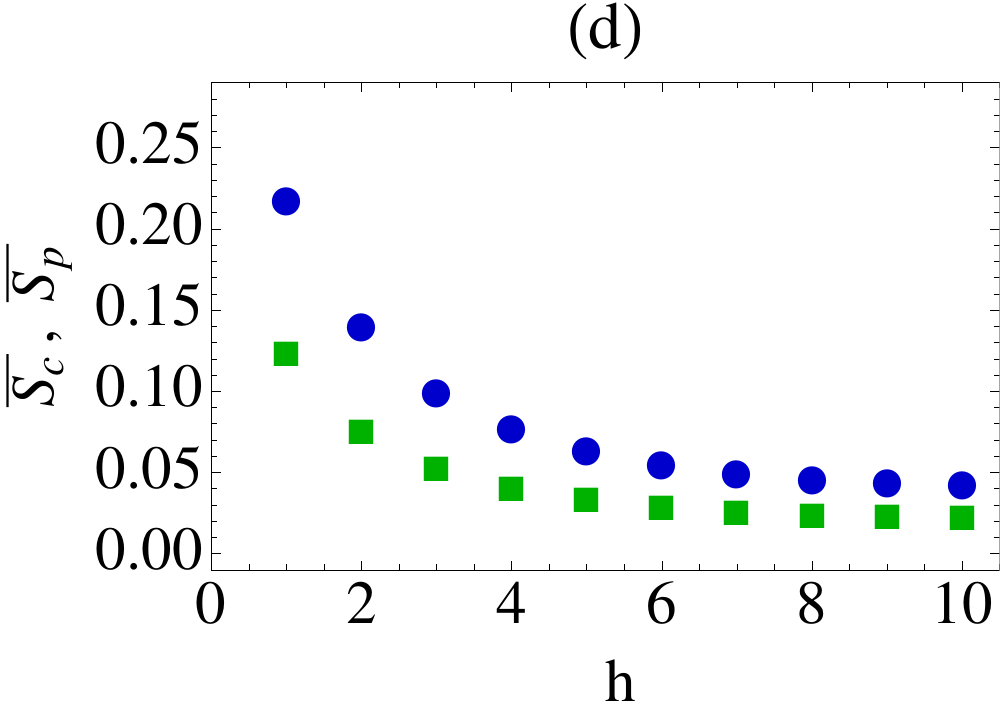}
\caption{(Color online) Subfigure {\bf (a)}: simulation of the complete (blue) and phase (green) synchronization measures~(\ref{PRIMA}) and  (\ref{Spq}) between the mechanical resonators as functions of time (in units of  $\tau=2 \pi / \omega_1$).
The dashed lines indicate the corresponding time averaged asymptotic values, {\it i.e.}\ the quantities 
$\bar{\mathcal S}_x=\lim_{T \rightarrow \infty}\frac{1}{T}\int_{0}^{T} \mathcal S_x(t) dt$ for $x=c,p$. 
Setting $\omega_1=1$ as a reference unit of frequency, the other physical parameters which have been used in the simulation are: $\omega_2=1.005$, $\gamma=0.005$, $\Delta_j=\omega_j$, $\kappa=0.15$, $g=0.005$, $\mu=0.02$, $n_b=0$ and $E=320$.
 Subfigure {\bf (b)}: 
time averaged complete (circles) and phase (squares) synchronization and Gaussian discord $D_G$ (diamonds) as functions of the coupling constant $\mu$. 
 Subfigure {\bf (c)}: time averaged synchronization measures as functions of the bath mean phonon number $n_b$. 
  Subfigure {\bf (d)}:  
 Synchronization between two arbitrary mechanical modes of a chain of $20$ coupled opto-mechanical systems
 as a function of the lattice distance $h$. All subsystems are assumed to have the same mechanical frequency $\omega=1$. }
 \label{plots}
\end{figure}

A simulation of the complete and phase synchronization between the mechanical modes is plotted in Fig.\ \ref{plots}.a using 
realistic values for the parameters \cite{review1,review2}  (see caption for details). 
After an initial transient, the system reaches a periodic steady state in which  $\mathcal S_c(t) $ and $\mathcal S_p (t)$
are significantly larger then zero, implying that 
both complete and phase synchronization take place in the system.  Their value is consistent
with the fundamental  limit  (\ref{heisenberg})  imposed by the Heisenberg principle and with the heuristic bound  (\ref{conjecture}) presented in the previous section. Indeed we numerically find that quantum squeezing
in the $p'_-(t)$ quadrature, needed to overcome the non-classicality threshold (\ref{bound2}), is absent in the system. 
Fig.\ \ref{plots}.b  and  Fig.\ \ref{plots}.c report instead  the behavior of the time averaged measures of complete and 
phase synchronization for different values of the coupling constant and of the bath temperature. We vary 
$\mu$ from zero \cite{mu} to a maximum threshold above which the classical equations are perturbed too much destroying the
limit cycles. 

Finally we have checked if quantum correlations are present in the system
verifying that, consistently with the  difference between entanglement and synchronization detailed in
the previous section, for many choices of the parameters  entanglement
negativity is always zero even though synchronization is reached. 
On the contrary,  non-zero level of Gaussian quantum discord \cite{adesso} (Fig.\ \ref{plots}.b) between the two mechanical modes is observed for all values of $\mu$ that lead to synchronization. 
Still our data are not sufficient to clarify  the functional relationship between discord and synchronization (if exists).
 
The synchronization observed between the oscillators is expected to  emerge also  when  more than two parties are present in the setup.
In particular we focus on the 
 case of   a (closed) chain formed by  $N$ opto-mechanical systems with first neighbor interactions (the  Hamiltonian being the natural generalization of~~(\ref{HAM})
 with uniform parameters).
 As before,  we enforce 
 the driving detuning condition 
$\Delta = \omega$ and set the laser intensities $E$ in order that each opto-mechanical system converges to a stable  limit-cycle.
Once these prerequisites are fulfilled we  linearize the dynamics around the classical steady state, which is assumed to be the same 
(synchronized) in each site, {\it i.e.}\
$\left< a_j (t) \right> = \alpha(t)$ and 
$\left< b_j(t) \right> = \beta(t)$ for all $j$.
This corresponds to a mean-field approximation applied {\it only} to the classical dynamics, while the fluctuation terms $a_j'$ and $b_j'$ 
can be treated exactly (without mean-field) since the associated Hamiltonian is quadratic. 
Fig.\ \ref{plots}.d reports the results obtained  for two mechanical modes separated by $h$ lattice steps:  
we notice  that the synchronization level among the various elements persists even if an  exponential decay in $h$ is present (a  behavior which is 
consistent with the one-dimensional topology induced by the selected  interactions).

{\it Summary: --} We have quantitatively studied the phenomenon of spontaneous synchronization in the setting of coupled CV quantum systems. We have shown that quantum mechanics sets universal limits to the level  of synchronization and discussed the relationship between this phenomenon and the emergence of quantum correlations. Finally we have analyzed the spontaneous synchronization of opto-mechanical arrays driven into self-sustained oscillations. A large number of open aspects are worth being  further investigated, among which: the interplay between quantum correlations and synchronization, the application of this theory to other physical systems like coupled optical cavities \cite{lee}, self-locking lasers \cite{laser}, {\it etc.}\ and the interpretation of synchronization as a useful resource for quantum communication and quantum control.

{\it Acknowledgments. --} This work has been supported by IP-SIQS, PRIN-MIUR and SNS (Giovani Ricercatori 2013). N.D. acknowledges support from CIFAR.

\newpage
\section{Supplemental Material: \\
Heuristic bound to phase synchronization}

The hypothesis underlying the conjecture~(7)   are:  (i) the system admits limit cycles which are approximately circular
$\langle a_j(t)\rangle \simeq r_j e^{-i \omega  t}$, (ii) thermal (or quantum) noise
is not phase sensitive in the sense that it is invariant for phase space rotations, and
(iii) the interaction potential between the two systems is of the
form $H_{int}= -\mu (a_1 a_2^\dag+a_2 a_1^\dag)$.  These assumptions are often valid for optical or mehcanical modes 
under the rotating wave approximation.
For such systems the interaction can be written in terms of symmetric and anti-symmetric
 normal modes $H_{int}= -\mu a_+^\dag a_+  + \mu a_-^\dag a_-$ which
 up to a renormalization of the bare frequencies ($\omega \rightarrow \omega - \mu $)
is equivalent to $H'_{int}=2 \mu a_-^\dag a_-$. This means that, because
of the interaction, the anti-symmetric mode rotates with a frequency $2\mu$ faster with respect to the symmetric mode.
 
 Since the phase of each individual limit cycle is arbitrary, the same must be true for the symmetric mode.
 Unless anti-synchronization appears, one can guess the following structure for the linearized equations for
 the fluctuations:
\begin{equation}
\frac{d}{dt}
\left[
\begin{array}{c}
q'_+  \\
p'_+ 
\end{array}
\right]
=
\left[
\begin{array}{cc}
- \gamma_{eff} &    0  \\
  0&    0  
\end{array}
\right]
\left[
\begin{array}{c}
q'_+  \\
p'_+  
\end{array}
\right]
+ \textrm{noise}. \label{symeq}
\end{equation}
The reason is that the symmetric mode must be stable in the amplitude (with effective
damping $\gamma_{eff}>0$) but its phase should be completely
 free to diffuse  (the corresponding Lyapunov exponent must be zero). 
 
 The dynamics of the anti-symmetric mode, linearized around a synchronized
 solution, will be like Eq.\ (\ref{symeq}) plus a frequency shift of $2\mu$ due to the interaction potential:
 \begin{equation}
\frac{d}{dt}
\left[
\begin{array}{c}
q'_-  \\
p'_- 
\end{array}
\right]
=
\left[
\begin{array}{cc}
- \gamma_{eff} &    -2 \mu  \\
  2\mu&    0  
\end{array}
\right]
\left[
\begin{array}{c}
q'_-  \\
p'_-  
\end{array}
\right]
+ \textrm{noise}. \label{antisymeq}
\end{equation}
The new matrix, for $\mu \neq 0$, has negative eigenvalues and this fact is the origin of synchronization.
Since we assumed the noise to be phase insensitive the diffusion matrix must be proportional to the identity.
In this case one can easily find the steady state and check that indeed $\langle p_{-}^{'2} \rangle \ge \langle q_{-}^{'2} \rangle$, and hence the bound (7) holds.
 
Of course this heuristic argument is very hand-waving but it gives the physical intuition that, in a classical
or  quantum system,  the precision of
phase synchronization may  be bounded by the precision of amplitude synchronization.


\begin{thebibliography}{99}


\bibitem{huygens} 
C.\ Huygens, {\it \OE{}uvres Compl\`etes de Christiaan Huygens}, vol. {\bf 15} pp.\ 243 (1893) Martinus
Nijhoff, The Hague; {\it ibid.} vol. {\bf 17} pp.\ 183 (1932).

\bibitem{csync} A.\ Pikovsky, M.\ Rosenblum and J.\ Kurths, {\it
Synchronization: A Universal Concept in Nonlinear Sciences}, Cambridge
University Press, New York  (2001).

\bibitem{CVSYSTEMS}
S. L. Braunstein and P. van Loock, Rev. Mod. Phys. 
{\bf 77}, 513 (2005); C. Weedbrook {\it et al.}, Rev. Mod. Phys. \textbf{84}, 621 (2012);
 A. Ferraro, S. Olivares and M. G. A. Paris, \emph{Gaussian
states in continuous variable quantum information}, (Bibliopolis, Napoli)
(2005).

\bibitem{review1} 
F.\ Marquardt, S.\ M.\ Girvin, {\it Physics} {\bf 2}, 40 (2009). 

\bibitem{review2}
M.\ Aspelmeyer, T.\ J.\ Kippenberg, F.\ Marquardt,
arXiv:1303.0733, (2013). 

\bibitem{protsyn}
V.\ Giovannetti, S.\ Lloyd, L.\ Maccone, J.\ H.\ Shapiro and F.\ N.\ C.\ Wong, Phys.\ Rev.\  A {\bf 70}, 043808 (2004);
R.\ Jozsa,  D.\ S.\ Abrams, J.\ P.\ Dowling, and C.\ P.\ Williams, Phys.\ Rev.\ Lett.\  {\bf 85}, 2010 (2000);
I.\ L.\ Chuang,  Phys.\ Rev.\ Lett.\ {\bf 85}, 2006 (2000);
V.\ Giovannetti, S.\ Lloyd, and L.\ Maccone,  Nature {\bf 412},   417-419 (2001).


\bibitem{quibit}
 O.\ V.\ Zhirov  and D.\ L.\ Shepelyansky, Phys.\ Rev.\ Lett.\ {\bf 100}, 014101 (2008); 
 S.-B.\ Shim,  M.\ Imboden and P.\ Mohanty, Science {\bf 316}, 95 (2007).


\bibitem{stochastic}
 I.\ Goychuk, J.\ Casado-Pascual, M.\ Morillo, J.\ Lehmann, and  P.\ H\"{a}nggi, Phys.\ Rev.\ Lett.\ {\bf 97}, 210601 (2006).


\bibitem{array4}
M.\ Ludwig and F.\ Marquardt, Eprint arXiv:1208.0327 [quant-ph] (2012). 


\bibitem{array5} U.\ Akram and G.\ Milburn, AIP Conf.\ Proc.\ {\bf 1363},
367 (2010). 

\bibitem{array6} A.\ Tomadin, S.\ Diehl, M.\ D.\ Lukin, P.\ Rabl and P.\ Zoller,
 Phys. Rev. A {\bf 86}, 033821 (2012).
 
 \bibitem{sync2} G.\ L.\ Giorgi, F.\ Galve, G.\ Manzano, P.\ Colet and R.\
Zambrini, Phys.\ Rev.\ A {\bf 85}, 052101 (2012);
G.\ Manzano, F.\ Galve, G.\ L.\ Giorgi, E.\ Hern\`andez-Garc\' ia and R.\ Zambrini,
Sci.\ Rep.\ {\bf 3}, 1439 (2013).

\bibitem{sync3} O.\ V.\ Zhirov and D.\ L.\ Shepelyansky,  Eur.\ Phys.\ J.\ 
{\bf D 38}, 375 (2006). 
\bibitem{sync4} S.\ Shim, M.\ Imboden, P.\ Mohanty, Science {\bf 316}, 95
(2007). 
\bibitem{sync5} M.\ Zhang {\it et al.}, Phys. Rev. Lett. {\bf 109}, 233906 (2012).

\bibitem{lee} T.\ E.\  Lee and M.\ C.\ Cross, arXiv:1209.0742, (2012).

\bibitem{matheny} M.\ H.\  Matheny {\it et al.}, arXiv:1305.0815, (2013).


\bibitem{array1} E.\ Buks and M.\ L.\ Roukes,  J. of Microel. Sys. {\bf
11}, 6 (2005). 

\bibitem{array7} Q.\ Lin {et al.}, Nature Photonics {\bf 4}, 236 (2010).
\bibitem{array8} M.\ Eichenfield, R.\ Camacho, J.\ Chan, K.\ J.\ Vahala and O.
Painter,  Nature {\bf 462}, 78 (2009). 
\bibitem{array9} D.\ E.\ Chang, A.\ H.\ Safavi-Naeini, M.\ Hafezi and O.\
Painter,  New J.\ Phys.\ {\bf 13}, 023003 (2011).



\bibitem{array2} G.\ Heinrich, M.\ Ludwig, J.\ Qian, B.\ Kubala and F.\
Marquardt, Phys.\ Rev.\ Lett.\ {\bf 107}, 043603 (2011).


\bibitem{factor} The normalization factor $\sqrt{2}$ is introduced for a  convenient notation in the quantization of the system. 

\bibitem{NOTE1}
 In a purely classical setting, $\langle \cdots\rangle$ of  Eq.~(\ref{PRIMA}) 
 accounts for taking the average over  several realization  of the stochastic process that tamper the system, or equivalently with respect to a time dependent phase space distribution. Notice that in this case, to put our definitions on a solid theoretical ground, 
 the variables $q_j(t),p_j(t)$ need to be not just 
 dimensionless (fundamental requirement to introduce $\varphi_j(t)$) but also  properly normalized in order to remove any ambiguity in the sum at the rhs of Eq.~(\ref{PRIMA}). For the models we are dealing with, i.e. systems of coupled harmonic oscillators, this can
be easily done by ensuring that when moving  into the quantum domain the observables associated with  $q_j$ and $p_j$ will allow us to express 
the local Hamiltonian as $H_j = \hbar \omega_j (p_j^2 + q^2_j)/2$ ($\omega_j$ being the corresponding frequencies). 

\bibitem{error}
If the averaged phase-space trajectories (limit cycles) of the two systems are constant but slightly different from each other, it means that this kind of error is not due to random noise but it is instead {\it systematic}. With the term {\it systematic} we mean that, with many measurements, this average  error can be estimated and subtracted from the measured data in order to single out the pure effect of quantum noise  on the amount of synchronization.

\bibitem{PHASE}
M. Ban, Phys. Lett. A {\bf 199} 275 (1995).

\bibitem{phase-diff} A.\  Luis and L.\ L.\ Sanchez-Soto, Phys.\ Rev.\ A {\bf 48}, 4702 (1993).


\bibitem{glauber} R.\ J.\ Glauber, Phys.\ Rev.\ {\bf 131}, 2766 (1963).


\bibitem{linearization}
D.\ Vitali, S.\ Mancini, L.\ Ribichini and P.\ Tombesi, Phys.\ Rev.\ A {\bf 65}, 063803 (2002);
A.\ Mari and J.\ Eisert, Phys. Rev. Lett. {\bf 103}, 213603 (2009);
 A.\ Farace and V.\ Giovannetti, Phys. Rev. A {\bf 86}, 013820 (2012). 

\bibitem{limit-cycles}
F.\ Marquardt, J.\ G.\ E.\ Harris and S.\ M.\ Girvin,\ Phys.\ Rev.\ Lett.\ {\bf 96}, 103901 (2006);
M.\ Ludwig, B.\ Kubala and F.\ Marquardt, New J.\ of Phys.\ {\bf 10}, 095013 (2008).

\bibitem{mu}
For $\mu=0$, the two resonators asymptotically acquire independent phases diffused along the respective limit cycles
of radius $R \simeq 500$ (with parameters of Fig.\ \ref{plots}). 
In this case we can estimate 
$\mathcal{S}_c= 2 \langle q_1^2+p_1^2+q_2^2+p_2^2 \rangle^{-1}\simeq R^{-2} \simeq 4 \times 10^{-5}$.

\bibitem{adesso}
G. Adesso and A. Datta, Phys. Rev. Lett. {\bf 105}, 030501 (2010).
\bibitem{EPR} A.\ Einstein, B.\ Podolsky, and N.\ Rosen, Phys.\ Rev.\ {\bf 47}, 777 (1935).
		


\bibitem{mixedsync1}  A.\ Prasad,  Chaos Solitons Fractals {\bf 43}, 42 (2010);
A.\ Sharma and M.\ D.\ Shrimali, Nonlinear Dyn. {\bf 69}, 371 (2012);
 K.\ Czolczynski, P.\ Perlikowski, A.\ Stefanski and T.\ Kapitaniak, Commun.\ Nonlinear Sci.\ Numer.\ Simulat.\ {\bf 17}, 3658 (2012).

\bibitem{P1} P.\ D.\ Drummond and C.\ W.\ Gardiner, J.\ Phys.\ A {\bf 13}, 2353 (1980).
\bibitem{P2} K.\ Dechoum {\it et al.}, Phys. Rev. A {\bf 70}, 053807 (2004).

\bibitem{laser} R.\ Graham, Springer Tracts in Mod.\ Phys.\ {\bf 66}, (1973).  

\end{thebibliography}
\end{document}